\documentclass[twocolumn,preprintnumbers,prl]{revtex4}
\usepackage{graphicx}
\usepackage{times}

\usepackage[hypertex]{hyperref}

\begin{document}

\title{How can we test seesaw experimentally?}

\author{Matthew R. Buckley and Hitoshi Murayama}
\affiliation{Department of Physics, University of California,
                Berkeley, CA 94720, USA}
\affiliation{Theoretical Physics Group, Lawrence Berkeley National Laboratory,
                Berkeley, CA 94720, USA}
\date{\today}

\begin{abstract}
  The seesaw mechanism for the small neutrino mass has been a popular
  paradigm, yet it has been believed that there is no way to test it
  experimentally.  We present a conceivable outcome from future
  experiments that would convince us of the seesaw mechanism.  It
  would involve a variety of data from LHC, ILC, cosmology,
  underground, and low-energy flavor violation experiments to
  establish the case.
\end{abstract}
\pacs{}
\maketitle

Recent years have seen revolutionary progress in neutrino physics
\cite{review}.  What used to be invisible particles Pauli regretted
proposing turned out to be a big excitement.  This is thanks to the
discovery of their quantum mechanical oscillation over macroscopic
distances, which implies they have tiny but finite masses against the
prediction of the Standard Model (SM) of particle physics.  Moreover,
neutrinos have relevance to many fields other than particle physics,
{\it e.g.}\/ nuclear physics, astrophysics, and cosmology.  They may
help explain why we exist at all (cosmic baryon asymmetry)
\cite{leptogenesis}, or why the universe is so big (inflation) if
combined with supersymmetry \cite{inflation}.

A possible finite neutrino mass has been of great interest to
physicists as a potential probe of physics at extremely high energies.
To parameterize physics at a high energy $\Lambda$, we can
systematically expand the Lagrangian in its inverse powers,
\begin{equation}
  {\cal L} = {\cal L}_{SM} + \frac{1}{\Lambda} {\cal L}_5 +
  \frac{1}{\Lambda^2} {\cal L}_6 + \cdots
\end{equation}
Here, ${\cal L}_{SM}$ is the Lagrangian of the SM which is
renormalizable and hence contains only operators of mass dimension
four or less.  Terms suppressed by inverse powers of $\Lambda$ are
non-renormalizable and represent the impact of physics at high
energies as suppressed effects at low energies that we may probe in
experiments.  Possible operators that may be present at each order in
$1/\Lambda$ can be enumerated with the particle content of the SM.
Even though there are a large number of possible operators in ${\cal
L}_6$ and beyond, there is only one operator one can write down in
${\cal L}_5$,
\begin{equation}
  {\cal L}_5 = \frac{1}{2} (LH) (LH).
\end{equation}
By substituting the vacuum expectation value for the Higgs field $v =
\langle H \rangle = 174$~GeV, this is nothing but the Majorana mass of
neutrinos,
\begin{equation}
  \frac{1}{\Lambda} {\cal L}_5 = \frac{1}{2} \frac{v^2}{\Lambda} \nu
  \nu = \frac{1}{2} m_\nu \nu \nu.
  \label{eq:D=5}
\end{equation}
Therefore, neutrino mass can be viewed as the leading order effect of
physics at high energies, and hence is very important.

The most striking aspect of the discovered neutrino masses is their
tininess.  Compared to masses of other elementary particles, neutrino
masses are seven or more orders of magnitude smaller.  Following the
above operator analysis, the smallness of neutrino mass $\sim 0.1$~eV
translates to extremely high energy scales $\Lambda \sim 10^{14}$~GeV.
This is an energy scale we cannot hope to reach directly with particle
accelerators.

In fact, it is incredibly fortunate that we could probe such tiny
neutrino masses at all.  Any kinematic effect of neutrino mass for a
typical accelerator or cosmogenic neutrino are suppressed by $(m_\nu /
E_\nu)^2 \sim (0.1~{\rm eV}/1~{\rm GeV})^2 = 10^{-20}$.  Even though
such tiny effects appear hopelessly small for experimental detection,
interferometry may help enhance their effects to observable size.
Interferometry requires three ingredients: a coherent source, the
presence of interfering waves, and long baselines.  Nature kindly
provided us all of these for neutrinos.  There are numerous coherent
sources of neutrinos, including the Sun, cosmic ray interactions in
the atmosphere, nuclear reactors and particle accelerators.  There are
interfering waves because of large mixing angles.  Lastly, there are
macroscopically long baselines available, such as the sizes of the
Earth or the Sun. An effect as small as $10^{-20}$ is observable
thanks to such fortuitous circumstances.

Having observed tiny neutrino masses, which could well be the impact
of physics at extremely high energies, we are posed with an obvious
challenge.  What is really going on at such high scales?  The standard
seesaw mechanism ~\cite{seesaw} introduces gauge-singlet right-handed
neutrinos to generate the operator Eq.~(\ref{eq:D=5}), but how will we
know if this is the case?  Without particle accelerators that reach
such high energies, it appears that our interpretation will remain
forever ambiguous.  One can write many theories that would give rise
to the observed tiny neutrino masses and mixings without any apparent
contradiction with all available data at energy scales we can directly
probe.  Is there a way to overcome this deadlock?  The conventional
answer is no; yet the opportunity to probe physics at such high
energies motivates us to seek for one.

In this Letter, we present a hypothetical yet conceivable outcome of
future experiments that would convince us of the physics responsible
for the tiny neutrino masses.  It would require the collection of many
different experimental approaches, including the Large Hadron Collider
(LHC), International Linear Collider (ILC), cosmology, underground,
and low-energy flavor violation experiments.  

We assume three important outcomes from the future experiments. First,
underground experiments establish the existence of neutrinoless double
beta decay of nuclei.  Second is the discovery of supersymmetry at the
LHC, followed up by the ILC verifying that it is indeed supersymmetry
\cite{FMPT} and measuring the masses of superparticle masses precisely
\cite{TFMYO}.  Third, we assume the measured masses, when extrapolated
to high energies, show unification at $M_{GUT} \sim 2 \times
10^{16}$~GeV, as already hinted by the precise measurements of gauge
coupling constants. Combined, such data would go a long way towards
establishing the seesaw mechanism.  Gaps in such a claim can be filled
in by other data such as the cosmic microwave background, large-scale
structure of the universe, and searches for low-energy lepton flavor
violation.

\begin{figure}[t]
  \centering
  \includegraphics[width=0.9\columnwidth]{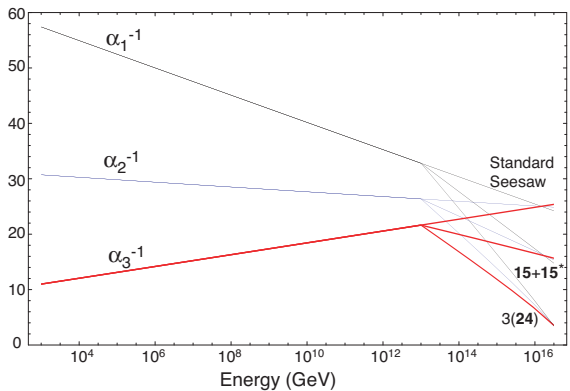}
  \caption{The apparent unification of gauge coupling constants with
    the two-loop renormalization group equation (RGE) with $m_{SUSY}
    \simeq 1$~TeV. Uncertainty in coupling constants from
    \cite{Eidelman:2004wy}.  Labels indicate particle content in
    addition to the MSSM at $M=10^{13}$~GeV, listed as representations
    of SU(5), and ``Standard Seesaw'' refers to gauge-singlet
    right-handed neutrinos.  }
  \label{fig:couplings}
\end{figure}

The unification of superparticle masses at $M_{GUT}$ is the crucial
aspect of the discussion.  Assuming the minimal supersymmetric
extension of the SM (MSSM), the gauge coupling constants apparently
unify at $M_{GUT}$, as shown in the ``Standard Seesaw'' bands in
Fig.~\ref{fig:couplings}.  However, this observation may be dismissed
as a pure coincidence.  Two lines are bound to meet at some energy
scale, while the third line meeting at the same point might happen
accidentally.

\begin{figure}[t]
  \centering
  \includegraphics[width=0.9\columnwidth]{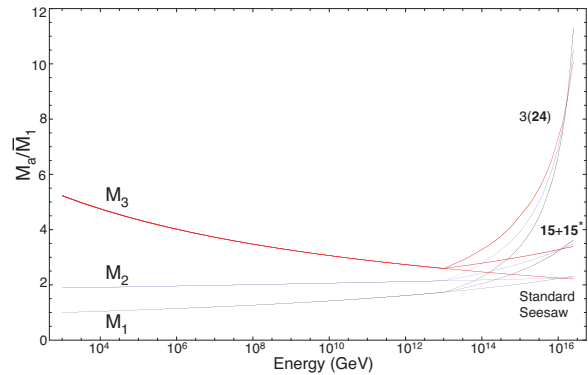}
  \caption{The unification of gaugino masses which data from LHC and
    ILC may demonstrate. One-loop RGE are used with $m_{SUSY} \simeq
    1$~TeV and $M=10^{13}$~GeV.  They are normalized by
    $\bar{M}_1\equiv M_1(1~\mbox{TeV})$.  The projected accuracy
    of measurements is from \cite{BPZ}. As in
    Fig.~\ref{fig:couplings}, labels indicate particle content in
    addition to MSSM. }
  \label{fig:gaugino}
\end{figure}

However, the measurement of superparticle masses will examine whether
the apparent unification is purely coincidental.  First, the masses of
the three gauginos, superpartners of SM gauge bosons, should unify at
the same energy scale $M_{GUT}$ if unification is true
\footnote{A notable exception is anomaly mediation: L.~Randall and R.~Sundrum, 
%``Out of this world supersymmetry breaking,''
Nucl.\ Phys.\ B {\bf 557}, 79 (1999);
% [arXiv:hep-th/9810155];
%%CITATION = HEP-TH 9810155;%% 
G.~F.~Giudice, M.~A.~Luty, H.~Murayama and R.~Rattazzi,
%``Gaugino mass without singlets,''
JHEP {\bf 9812}, 027 (1998).
% [arXiv:hep-ph/9810442].
%%CITATION = HEP-PH 9810442;%% 
}.  Unification of two masses would present a non-trivial test, while
the unification of the third constitutes another.  Such an observation
would therefore add two more non-trivial coincidences.  It has been
shown by simulations that the combination of LHC and ILC data can
provide sufficiently precise gaugino masses and hence non-trivial
tests of gaugino mass unification (Fig.~\ref{fig:gaugino})
\cite{TFMYO,BPZ}.

In addition, the masses of matter superpartners (sfermions) can also
be measured precisely.  If they exhibit the unification at $M_{GUT}$,
then grand unification would be very difficult to dismiss.  The
left-handed quarks $Q$, right-handed up-quarks $U$ and right-handed
charged leptons $E$ of a given generation would belong to the same
multiplet, and their superpartner masses unify at $M_{GUT}$.  If this
happens for all three generations, it would present six more
coincidences.  Unification of the right-handed down quarks $D$ and
left-handed leptons $L$ would provide three more coincidences.

\begin{figure}[t]
  \centering
  \includegraphics[width=0.9\columnwidth]{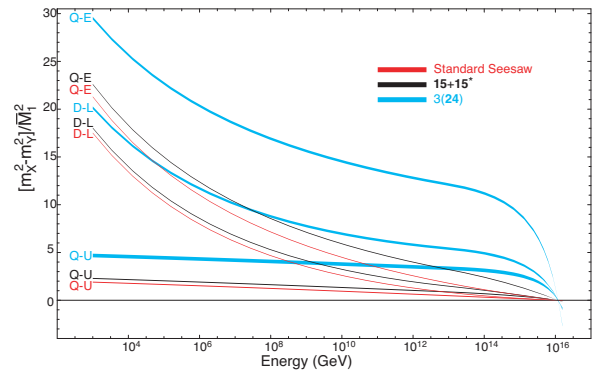}
  \caption{Prediction on sfermion masses assuming their unification at
    the same $M_{GUT}$ suggested by the gauge couplings and gaugino
    masses for three different particle contents at $M=10^{13}$~GeV.
    $\bar{M}_1\equiv M_1(1~\mbox{TeV})$, and $m_X^2-m_Y^2$ refer
    to mass squared differences between scalar $Q$ and $U$, $Q$ and
    $E$, or $D$ and $L$.}
  \label{fig:scalar} 
\end{figure}

The main point is that the combination of gaugino mass and sfermion
mass unification provides important information about the particle
content between TeV and $M_{GUT}$ \cite{KMY}.  For instance, gaugino
mass unification holds even if there are multiple stages of symmetry
breaking as long as they are eventually unified in a single group.  In
addition, we see an apparent gaugino mass unification in models of
gauge mediation with messengers that happen to fall into complete
SU(5) multiplets \cite{DNS} even if there is no true unification.  On
the other hand, these different possibilities give different patterns
of sfermion masses and can be discriminated against.  If the simple
picture of unification as in the ``Standard Seesaw'' of
Fig.~\ref{fig:scalar} holds, we know it is not gauge mediation, and
there is no additional stage of symmetry breaking.  To the extent that
we do not dismiss so many non-trivial consistency checks of
unification as mere coincidences, the particle content between TeV and
$M_{GUT}$ would be subject to stringent constraints; namely that there
cannot be any new particles with non-trivial quantum numbers under the
SM gauge groups.

Such an observation would strongly favor the possibility that the
neutrino masses originate from gauge-singlet particles, that is, the
standard seesaw mechanism.  Note that we already know the energy scale
responsible for neutrino masses is substantially lower than $M_{GUT}$.
The largest neutrino mass cannot be smaller than $(\Delta
m^2_{23})^{1/2} \simeq 0.05$~eV, which implies $\Lambda \lesssim 6
\times 10^{14}~{\rm GeV} \ll M_{GUT}$.  Additional particles needed at
or below $\Lambda$ cannot have non-trivial SM charges.  We will
discuss this constraint more quantitatively below.

First, we need to know that the neutrino mass is given by the operator
Eq.~(\ref{eq:D=5}).  This is where underground experiments searching
for neutrinoless double beta decay ($0\nu\beta\beta$) of nuclei come
in.  Once a positive signal is established, we would conclude that
neutrinos are Majorana particles and hence the operator
Eq.~(\ref{eq:D=5}) exists.  In addition, the rate would also determine
the energy scale $\Lambda$.  The rate of $0\nu\beta\beta$ determines
the effective electron neutrino mass (up to uncertainties in nuclear
matrix elements) $ \langle m_\nu \rangle_{ee} \equiv \left| \sum_i
m_{\nu_i} U_{ei}^2 \right|$.  For example
$\langle m_\nu\rangle _{ee} \simeq 0.1$~eV would translate to 
$\Lambda \simeq 3 \times 10^{14}$~GeV, substantially lower than 
$M_{GUT}$.  Then the question is what set of particles would produce 
the operator Eq.~(\ref{eq:D=5}) with this value of $\Lambda$.

\begin{figure}[t]
  \centering
  \includegraphics[width=0.9\columnwidth]{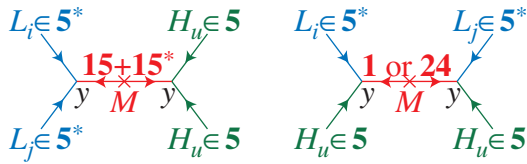}
  \caption{Possible supergraphs which generate neutrino mass consistent with
    gauge coupling and gaugino mass unification.}
  \label{fig:seesaw}
\end{figure}

Second, the constraint from the gauge coupling and gaugino mass
unification is that any additional particles below $M_{GUT}$ must
appear in complete SU(5) multiplets.  Therefore there are only a
finite number of possibilities to generate the operator
Eq.~(\ref{eq:D=5}).  Because of supersymmetry, the operator needs to
be in a superpotential and, due to the non-renormalization theorem of
the superpotential, can only be generated by a tree-level exchange of
new particles.  These can be either in the $LL$ to $HH$ channel, or in
the $LH$ to $LH$ channel (Fig.~\ref{fig:seesaw}).  Because we know
already that at least two of the neutrinos have finite yet different
masses, the $LL$ channel must be in the symmetric combination of
flavors.  Since $L$ belongs to ${\bf 5}^*$ multiplet in SU(5), the
symmetric combination of two ${\bf 5}^*$ can only be ${\bf 15}^*$.  We
also need a ${\bf 15}$ multiplet to avoid anomalies and allow for its
mass.  For the $LH$ to $LH$ channel, where $H$ belongs to the ${\bf
5}$, the exchanged particle can be either in ${\bf 24}$ or ${\bf 1}$.
In order to generate a neutrino mass matrix of rank $\geq 2$, we need
at least two copies of ${\bf 24}$ or ${\bf 1}$.  For quantitative
analysis, we assume all three neutrinos have mass and hence three
copies of ${\bf 24}$ or ${\bf 1}$.  Therefore, there are three logical
possibilities to be studied: the standard seesaw with three ${\bf 1}$,
the modified seesaw with three ${\bf 24}$, and the so-called Type-II
seesaw with ${\bf 15}+{\bf 15}^*$.

The effects of these extra multiplets on the running of coupling
constants and gaugino masses are demonstrated in
Figs.~\ref{fig:couplings} and ~\ref{fig:gaugino} for the particular
choice $M=10^{13}$~GeV (see {\it e.g.}\/~\cite{MVBBO} for RGEs). As
previously noted, unification at $M_{GUT} \sim 2\times 10^{16}$~GeV
remains unchanged.  Note that the scale $\Lambda$ is not the same as
the mass of these additional multiplets $M$ because their relationship
depends on the size of the Yukawa couplings $y$ as $\Lambda = M/y^2$.
Imposing the perturbativity of the Yukawa couplings to be consistent
with perturbative unification, $y \lesssim O(1)$ and hence $M \lesssim
\Lambda \ll M_{GUT}$.

Third, these additional particles below $M_{GUT}$ would affect the
evolution of sfermion masses.  The presence of additional particles
cause larger gaugino masses above the scale $M$ (as in
Fig.~\ref{fig:gaugino}), and hence larger RGE effects in sfermion
masses-squared which are proportional to gaugino masses-squared.  One
can then discuss the mass-squared differences of matter superpartners
in the same SU(5) multiplets in the unit of gaugino masses.  We use
the ratios $(m^2_Q - m^2_U)/\bar{M}_1^2$, $(m^2_Q -
m^2_E)/\bar{M}_1^2$, and $(m^2_D - m^2_L)/\bar{M}_1^2$ with $\bar{M}_1
= M_1({\rm TeV})$ for this purpose.  At the leading order in RGE with
negligible Yukawa couplings, these quantities are independent of the
boundary conditions and hence allow for definite predictions.
Therefore, we restrict our quantitative analysis to the leading order
({\it i.e.}\/~one-loop), yet we stick to two-loop RGE for gauge
coupling constants to be consistent with $M_{GUT} = 2\times
10^{16}$~GeV.  Higher order RGE for sfermion masses will not change
the results qualitatively \cite{preparation}.  For the three logical
possibilities consistent with gauge coupling and gaugino mass
unification, we find different values for these ratios as seen in
Fig.~\ref{fig:scalar} and Table~\ref{tab:ratios}.  The quantitative
result obviously depends on $M$.  The main conclusion is that the
three different models can be distinguished from each other if percent
level measurements on these mass ratios can be performed at the LHC
and ILC and if $M \lesssim 10^{14}$~GeV.

\begin{table*}[t]
  \centering
  \begin{tabular}{|c|c|c|c|c|c|c|c|}
    \hline
    $M$ & & \multicolumn{2}{c|}{$10^{15}$~GeV} &
    \multicolumn{2}{c|}{$10^{14}$~GeV} &
    \multicolumn{2}{c|}{$10^{13}$~GeV} \\ \hline
    model & MSSM & $3\times{\bf 24}$ & ${\bf 15}+{\bf 15}^*$ &
    $3\times{\bf 24}$ & ${\bf 15}+{\bf 15}^*$ & $3\times{\bf 24}$ &
    ${\bf 15}+{\bf 15}^*$ \\ \hline 
    $(m_Q^2-m_U^2)/M_1^2$ & $1.90^{+0.05}_{-0.05}$ &
    $1.98^{+0.05}_{-0.05}$ & $1.93^{+0.05}_{-0.05}$ &
    $2.41^{+0.05}_{-0.06}$ & $2.04^{+0.05}_{-0.05}$ &     
    $4.68^{+0.19}_{-0.19}$ & $2.29^{+0.05}_{-0.05}$ \\ \hline
    $(m_Q^2-m_E^2)/M_1^2$ & $21.30^{+0.03}_{-0.04}$ &
    $21.41^{+0.03}_{-0.04}$ & $21.35^{+0.03}_{-0.04}$ &
    $22.58^{+0.04}_{-0.04}$ & $21.70^{+0.04}_{-0.04}$ &
    $29.52^{+0.14}_{-0.13}$ & $22.60^{+0.04}_{-0.04}$\\ \hline
    $(m_D^2-m_L^2)/M_1^2$ & $17.48^{+0.05}_{-0.03}$ &
    $17.50^{+0.03}_{-0.04}$ & $17.49^{+0.03}_{-0.03}$ &
    $17.77^{+0.04}_{-0.04}$ & $17.62^{+0.04}_{-0.04}$ &
    $20.15^{+0.13}_{-0.13}$ & $18.02^{+0.04}_{-0.04}$ \\ \hline
  \end{tabular}
  \caption{The predicted mass ratios at 1~TeV for three different possible
    origins of neutrino mass consistent with gauge coupling and
    gaugino mass unification, for three values of heavy 
    particle mass $M$.  The errors are due to uncertainties in the observed
    gauge coupling constants and expected experimental uncertainties 
    in the gaugino mass.}
  \label{tab:ratios}
\end{table*}

Can such precise measurements be done?  According to the studies, the
slepton and gaugino masses can be measured at permille levels at the
ILC, negligible errors for our purpose.  The question is the
measurement of squark masses.  The LHC can achieve statistical
accuracy of 0.2\% on measurements of squark mass, yet is limited by
the systematic uncertainty in the jet energy scale expected at the 1\%
level \cite{LHC/LC}.  At the ILC, kinematic distribution in the squark
decay product would give better than 1\% measurement if enough
luminosity is obtained and jet energy calibrated by $Z$ mass
\cite{FF}.  In addition, a threshold scan would possibly lead to a
$\sim 0.5\%$ level measurement \cite{Blair}.  ILC is crucial in this
program because we have to differentiate different types of squarks.
The required precision is challenge even for the ILC, yet it is
encouraging that the measurement strategies have not yet been fully
optimized.

Therefore, it is quite conceivable that LHC and ILC measurements of
superparticle masses would pick one out of three possibilities for
additional particle content.  In particular, observation of sfermion
mass unification with the MSSM particle content would mostly likely
convince us that unification is real, and hence exclude the additional
gauge non-singlet particles below $M_{GUT}$ as in the modified Type-I
or Type-II models.  By process of elimination, it would establish the
SM singlets as the origin of neutrino masses, and hence the standard
seesaw mechanism.

This result would also tell us something about the origin of baryon
asymmetry.  Note that the tremendous success of the inflationary
paradigm suggests the baryon asymmetry must be generated by physics at
or below the inflationary scale $H_{\it inf}$.  The current
cosmological data provide an upper limit $H_{\it inf} \leq 1.5 \times
10^{14}$~GeV \cite{Martin:2006rs}.  Therefore, in generic inflationary
models \footnote{It may be possible to evade this constraint in models
  of inflation with multiple energy scales and/or preheating that
  produce particles heavier than $H_{\it inf}$ after inflation.},
baryogenesis would require particles $\lesssim 10^{14}$~GeV, and hence
they would affect the scalar mass unification if not gauge singlets.
Once scalar mass unification is confirmed, baryogenesis must be either
due to particles in the MSSM, namely electroweak baryogenesis, or due
to gauge-singlets beyond the MSSM.  The former can in principle be
excluded experimentally ({\it e.g.}\/, searches for light $\tilde{t}$
and $\chi^{\pm}$ and electric dipole moments \cite{Balazs:2004ae} and
$B$ physics \cite{Murayama:2002xk}).  This would then require
baryogenesis by gauge-singlets, hinting very strongly at leptogenesis
\cite{leptogenesis}.  In a similar fashion, we could obtain
interesting restrictions on various axion models which also require
additional gauge non-singlet particles $\lesssim 10^{12}$~GeV.

We have to mention potential loopholes with the whole argument.
First, the observation of $0\nu\beta\beta$ would not necessarily
establish the Majorana neutrino mass as the dominant contribution.  It
could be, for example, due to $R$-parity violating supersymmetry or
extended gauge sector.  $R$-parity violation can be excluded if the
collider measurements of superparticle spectrum and couplings gives
the cosmic abundance of the lightest supersymmetric particle
consistent with the cosmological data (see, {\it e.g.}\/,
\cite{BBPW}).  Extended gauge sectors necessarily require additional
particles below $M_{GUT}$ which are also excluded by sfermion mass
unification.  Second, running of sfermion masses may also be affected
by the neutrino Yukawa couplings $y$ if they are close to $O(1)$.
However, such large $y$ tend to lead to sizable lepton-flavor
violation.  In fact, the present upper limit on $\mu \rightarrow
e\gamma$ combined with the now-established large mixing angle solution
to the solar neutrino problem already requires small $y$, and hence
favors $M \lesssim 10^{13}$~GeV for moderate values of supersymmetric
parameters \cite{HN}.  Future improvements on upper limits on such
processes combined with supersymmetric parameters from colliders would
yield stronger upper limit on the mass $M$ which can only strengthens
the result.

We have presented a hypothetical yet conceivable outcome from future
experiments that would establish the standard seesaw mechanism.  It is
surprising that collider measurements of superparticle masses would be
the crucial information, while additional information from
$0\nu\beta\beta$, cosmological abundance of dark matter, and
low-energy lepton flavor violation would fill in the gaps.  This way,
we may use the neutrino masses as real probe to physics at extremely
high energy scales.

\begin{acknowledgments}
  This work was supported in part by the U.S. DOE
  under Contract DE-AC03-76SF00098, and in part by the NSF
  under grants PHY-00-98840 and PHY-04-57315.
\end{acknowledgments}

\end{document}